\documentclass[aps,prl,reprint,showpacs,twocolumn,superscriptaddress,floatfix]{revtex4-1}
\usepackage{graphicx}
\usepackage{amssymb}
\usepackage{amsmath}

\begin{document}

\title{"Smile"-Gap in the Density of States of a Cavity between Superconductors}

\author{J. Reutlinger}

\affiliation{Fachbereich Physik, Universit\"at Konstanz, D-78457 Konstanz, Germany}

\author{L. Glazman}

\affiliation{Department of Physics, Yale University, New Haven Connecticut 06511-8499, USA}

\author{Yu. V. Nazarov}

\affiliation{Kavli Institute of Nanoscience Delft, Delft University of Technology, 2628 CJ Delft, Netherlands}

\author{W. Belzig}

\email[]{Wolfgang.Belzig@uni-konstanz.de}

\affiliation{Fachbereich Physik, Universit\"at Konstanz, D-78457 Konstanz, Germany}



\date{\today}

\begin{abstract}
The density of Andreev levels in a normal metal (\textit{N}) in contact with two superconductors (\textit{S}) is known to exhibit an induced minigap related to the inverse dwell time. We predict a small secondary gap just below the superconducting gap edge---a feature that has been overlooked so far in numerous microscopic studies of the density of states in \textit{S-N-S} structures. In a generic structure with \textit{N} being a chaotic cavity, the secondary gap is the widest at zero phase bias. It closes at some finite phase bias, forming the shape of a "smile". Asymmetric couplings give even richer gap structures near the phase difference $\pi$. All the features found should be amendable to experimental detection in high-resolution low-temperature tunneling spectroscopy.
\end{abstract}


\maketitle

The modification of the density of states (DOS) in a normal metal by a superconductor in its proximity was discovered almost 50 years ago \cite{andreev:64}. Soon afterwards, it was predicted, theoretically, for diffusive structures that a so-called minigap of the order of the inverse dwell time in the normal metal (or the Thouless energy) appears in the spectrum \cite{mcmillan:68}. 
The energy-dependent DOS reflects the energy scale of electron-hole decoherence, and is sensitive to the distance, the geometry and the properties of the contact between the normal metal and the superconductor \cite{golubov:88,beenakker:92,belzig:96}. The details of the local density of states in proximity structures have been investigated  experimentally many years later \cite{gueron:96,scheer:01,moussy:07,mourik:12,churchill:13} and the theoretical predictions  have been confirmed \cite{leseur:08,pillet:10,wolz:11} in detail.

Substantial interest has been paid to the density of states in a finite normal metal between two superconducting leads with different superconducting phases \cite{belzig:99,golubov:04}. The difference between diffusive \cite{belzig:96} and classical ballistic \cite{lodder:98} dynamics has been investigated \cite{pilgram:00}. Many publications have addressed the dependence of the minigap on the competition between dwell and Ehrenfest time \cite{VavilovLarkin:03, beenakker:05}. The most generic model in this context is that of a {\it chaotic cavity}, where a piece of normal metal is connected to the superconductors by means of  ballistic point contacts that dominate the resistance of the structure in the normal state. The Thouless energy is given by  $E_{\textrm{Th}} = (G_\Sigma/G_Q) \delta_S$, $G_Q = e^2/\pi\hbar$ being the conductance quantum, $G_\Sigma \gg G_Q$ being the total conductance of the contacts and $\delta_S$ being the level spacing in the normal metal provided the contacts are closed. The DOS in chaotic cavities has been studied for years. \cite{melsen:97,beenakker:05,kuipers:10,kuipers:11} 

\begin{figure}[t] 
 \includegraphics[width=0.9\columnwidth,angle=0]{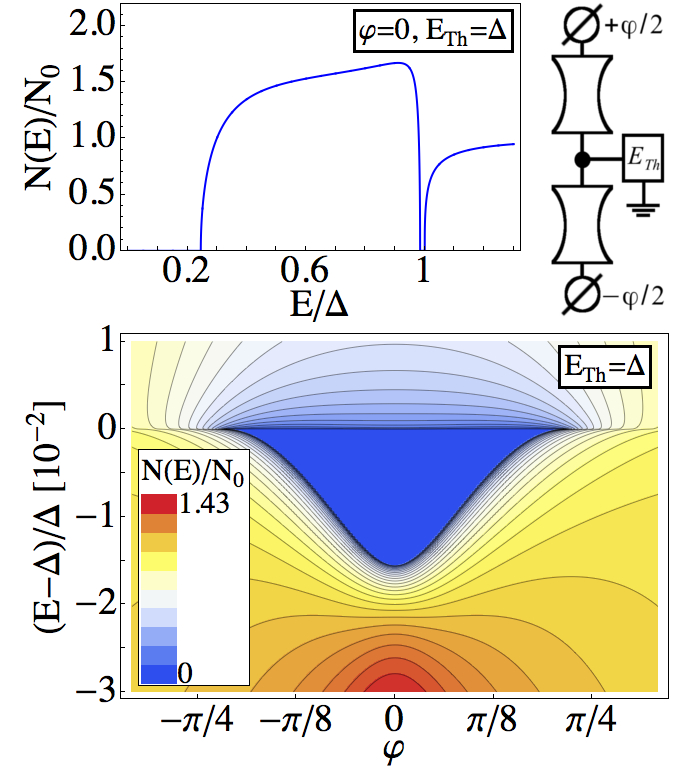}
\caption{\label{fig:setup}
Upper plot: DOS in the central region at zero phase difference
and $E_{\textit{Th}}=\Delta$ showing the usual minigap around $E = 0$ and
additionally a secondary gap below $E = \Delta$.
On the right: Quantum circuit theory [26] diagram of the system under investigation. 
A pseudoterminal labeled with $E_{\textrm{Th}}$ accounts for random phase shifts between electron and hole components of the quasiparticle wave functions  (not implying an electric connection to the ground)  
Lower plot: DOS near  $\Delta$ illustrating the phase dependence of the secondary gap.
}
\end{figure}

The DOS depends on the ratio of $E_{\textrm{Th}}$ and the superconducting energy gap $\Delta$, and on the superconducting phase difference. If the dwell time exceeds the Ehrenfest time, qualitative features of the DOS do not seem to depend much on the contact nature and are the same for ballistic, diffusive and tunnel contacts. Mesoscopic fluctuations of the DOS \cite{taras:01,ostrovsky:01} are small provided  $G\gg G_Q$. It looks like everything is understood, perhaps  except a small dip or peak in the DOS just at the gap edge for the diffusive case, which has been seen in \cite{golubov:88,belzig:96,golubov:96,bezuglyi:03,levchenko:08,bezuglyi:11}, but never attracted proper attention. 


In this Letter we demonstrate that the appearance of a secondary gap, in addition to the well-known gap $\sim E_{\textrm{Th}}$ in the DOS around the Fermi level, is a generic feature of \textit{S-N-S} structure containing high-transmission \textit{S-N} contacts. In the case of chaotic cavity with two identical ballistic contacts, the secondary gap exists for any $E_{\textit{Th}}>0.68\Delta$ and vanishes as $\Delta/E_{\textit{Th}}^2$ for $E_{\textit{Th}}\gg\Delta$, where $\Delta$ is the superconducting gap.
The gap closes at finite superconducting phase difference $\varphi$ and its contour in the energy-phase plot forms a characteristic smile pattern, see Fig.~\ref{fig:setup}. 
The gap can be attached to the gap edge at $E=\Delta$ at one of the boundaries or completely detached from it. For asymmetric contacts, more secondary gaps can emerge near a phase difference $\varphi=\pi$ and close to zero energy, see Fig.~\ref{fig:asymm}. 
The developed theory is applicable, for example, to an \textit{S-N-S} system with \textit{S-N} junctions of area $s\ll L^2$ and with the linear dimension $L$ of the normal region sufficiently small, $L/\xi\lesssim s/L^2$, compared to the coherence length $\xi=\hbar v_F/\Delta$ of the superconductors.



In a general context, we have demonstrated an emergence of the gaps in the semiclassical spectrum of a random quantum system upon changing a parameter.
In this spectrum, the neighboring levels are typically separated by the small distance $\simeq \delta_S$: the change of parameter causing the gap opening will split the system of levels pushing them apart at a much larger energy scale. This opens up interesting possibilities for quantum manipulation. The statistics of random levels that has been elaborated in detail \cite{altshuler:96} for the case of uniform semiclassical density of states is a challenge to understand in the case of such emergent gaps. Such understanding is likely to advance the theory of quantum disordered systems in general.

\begin{figure}[t] 
 \includegraphics[width=0.9\columnwidth,angle=0,clip=true]{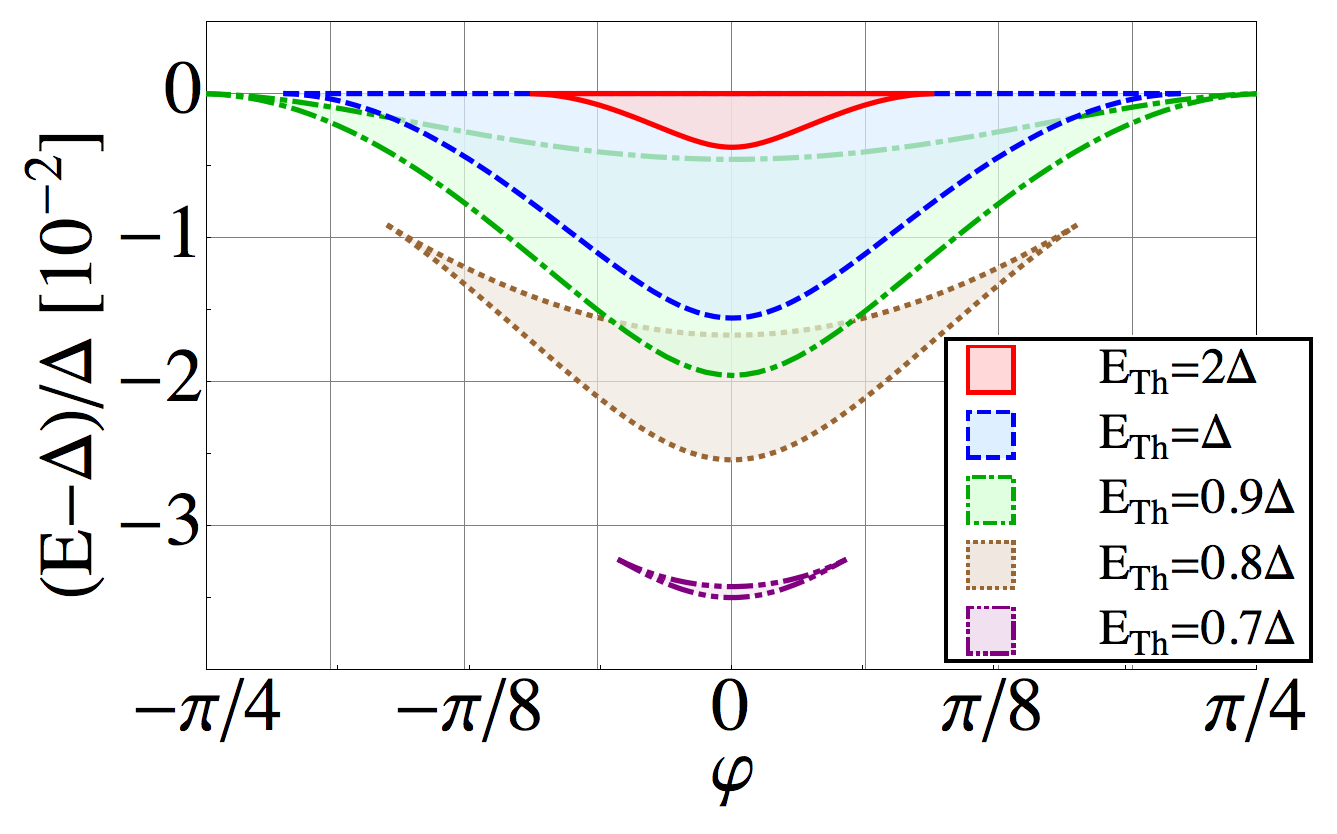}
 \includegraphics[width=0.95\columnwidth,angle=0,clip=true]{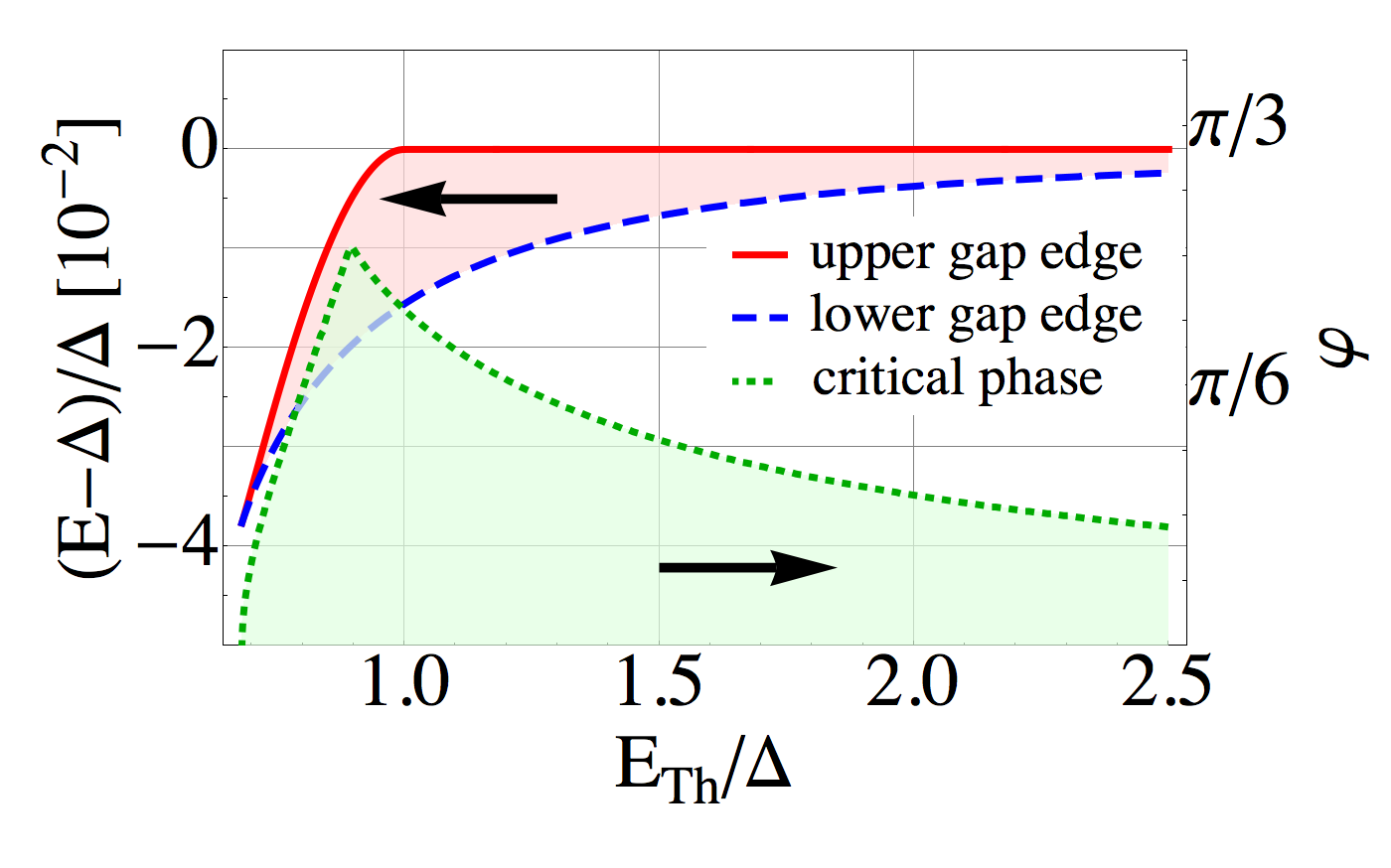}
\caption{\label{fig:gaps}
Upper plot:
Disappearing secondary gap for $E_{\textrm{Th}}<\Delta$. The colored regions represent the gapped part of the spectrum for different Thouless energies.
Lower plot:
Critical parameters of the secondary gap for $G_1=G_2$ in dependence of $E_{\textrm{Th}}$: Upper gap edge for $\varphi=0$ (red), lower gap edge for $\varphi=0$ (dashed blue) and critical phase (dotted green). The colored regions denote the gap.}
\end{figure}

To find the density of states in the cavity between two superconducting leads, we make use of the powerful quasiclassical Green's function method. Unlike the bridge geometry of \cite{levchenko:08}, we assume that, for the geometry we consider the spatial dependence of the Green's functions is not relevant. Therefore, we can make use of the discretized form of the method---the so-called quantum circuit theory \cite{nazarov:94,qt}. 
The crucial equation that relates the retarded matrix Green's functions $\hat G_c$
in the normal metal and those $\hat{G}_{1,2}$ in the two superconductors $1,2$, takes the form of matrix current conservation
\begin{equation}
	\label{eq:curr_cons}
 	\hat I_{1c}+\hat I_{2c} +iG_{\Sigma}(E/E_{Th})[\hat\tau_3,\hat G_c(E)]=0\,.
\end{equation}
Here, the matrix currents for ballistic junctions are given by $\hat I_{ic}=G_i [\hat G_c,\hat G_i]/(1+\{\hat G_c,\hat G_i\}/2)$ where $G_{1,2}$ are the conductances of the contacts  \cite{nazarov:99} and $G_{\Sigma}=G_1+G_2$ is the total conductance. $\hat\tau_i$ denote the Pauli matrices in Nambu space and $E_{\textrm{Th}}$ describes electron-hole decoherence in the normal metal due to a finite dwell time. To avoid confusion, we remark that electron-hole decoherence here refers to the effect of randomization of the relative phase between electron and hole. The two superconducting leads have the same energy gap $\Delta$ and phases $\pm\varphi/2$, so that the corresponding Green's functions read 
$\hat G_{1,2}=c\hat\tau_3+is[\hat\tau_1\cos(\varphi/2)\pm\hat\tau_2\sin(\varphi/2)]$
with the spectral functions $c$ and $s$ being given by 
$c=\sqrt{1+s^2}=E/\sqrt{E^2-\Delta^2}$ 
for $E>\Delta$ and by 
$c=\sqrt{1+s^2}=-iE/\sqrt{\Delta^2-E^2}$ 
for $E<\Delta$. 
The Green's function in the normal metal node is parametrized as 
$\hat G_c=g\hat\tau_3+if[\hat\tau_1\cos(\phi/2)-\hat\tau_2\sin(\phi/2)]$.
Finally, Eq.~(\ref{eq:curr_cons}) has to be solved under the constraint $\hat G_c^2=1$, which is equivalent to $g^2-f^2=1$.  In the general case, one has to find a numerical solution of two equations in two complex variables.  The density of states $N(E)$ is finally obtained from $\hat{G}_c$ by using $N(E)/N_0=\mathrm{Re}\{\mathrm{Tr}\hat \tau_3\hat G_c(E)\}/2=\mathrm{Re}\{g\}$, $N_0$ being the density of states in the normal case.

First, we discuss the situation for a symmetric setup $G_1=G_2$.
In this case, the phase of the central node is determined from symmetry as $\phi=0$ and the problem is reduced to solving a single-variable equation. 
We transform 
Eq.~(\ref{eq:curr_cons}) into
\begin{equation}
	\label{eq:curr_cons2}
	i\frac{E}{E_{\textrm{Th}}}f + \frac{g s \cos(\varphi/2)-f c}{1+c g-s f\cos(\varphi/2)}=0\,.
\end{equation}
After elimination of $f$, Eq.~(\ref{eq:curr_cons2}) is solved numerically. The resulting density of states showing the secondary gap and its generic properties for different Thouless energies and phases are summarized in Figs.~\ref{fig:setup} and \ref{fig:gaps}. 

\begin{figure}[t] 
\vspace{5mm}
 \includegraphics[width=0.8\columnwidth,angle=0]{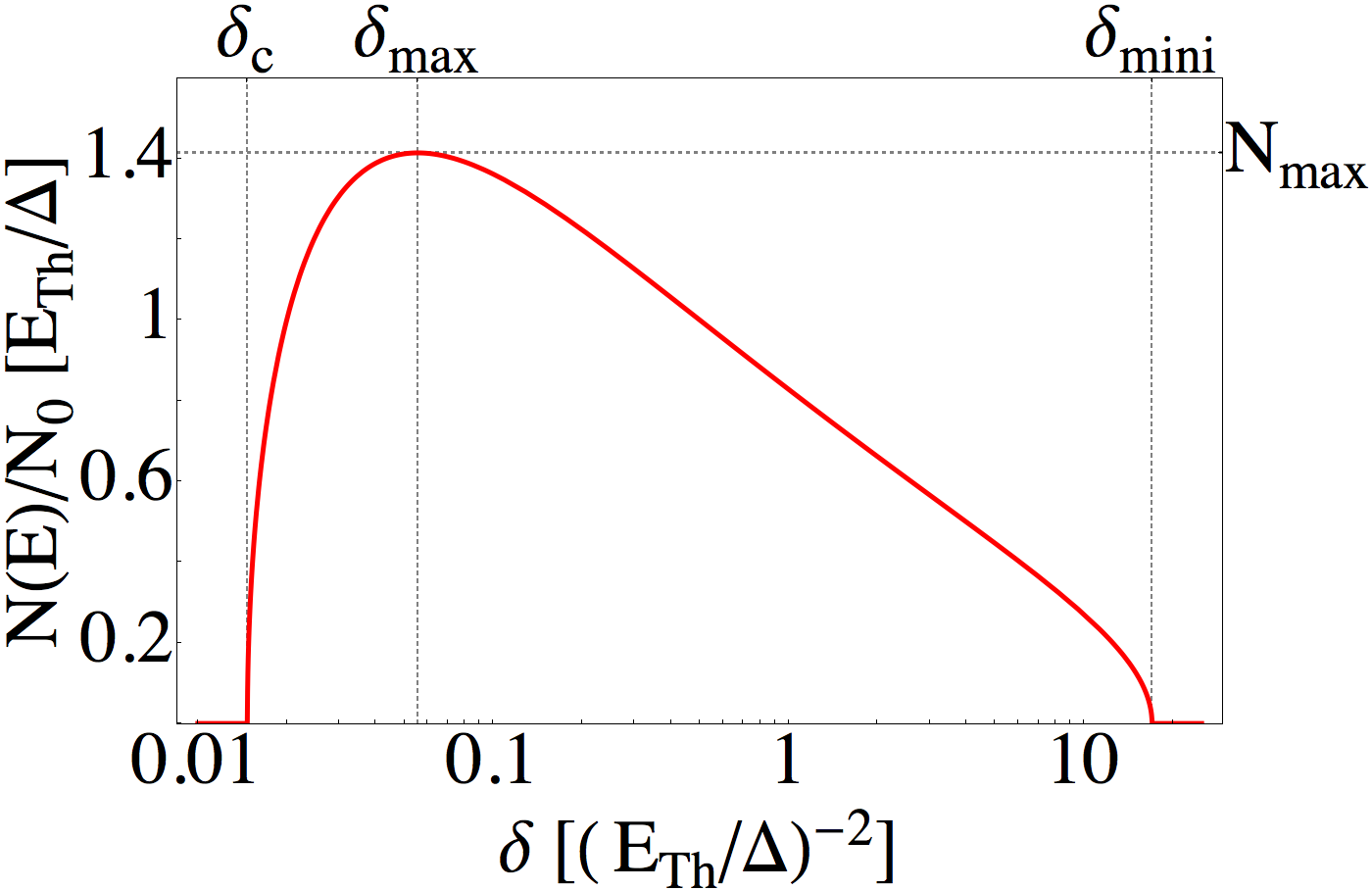}
\caption{\label{fig:universal}
Universal shape of the DOS for $\varphi=0$ between the gaps in the limit $E_{\textrm{Th}}\gg\Delta$. The curve is obtained from Eq.~(\ref{eq:curr_cons3}) and the characteristics are given in the main text.}
\end{figure}

Before discussing the full numerical solution, we present analytical results in the  limit of large Thouless energies $E_{\textrm{Th}} \gg \Delta$. 
We linearize Eq.~(\ref{eq:curr_cons2}) in the parameter range of interest (i.e., $ \delta=\left( \Delta - E \right)/\Delta\ll 1, \varphi \ll 1$ ) and  find
\begin{equation}
	\label{eq:curr_cons3}
	\frac{1}{2g^2}-\left( \frac{\varphi^2}{8}-\delta \right)+
	\frac{\Delta}{E_{\textrm{Th}}} \left[\frac{i}{2 g}+
	i\left(\frac{\varphi^2}{8}-\delta \right)g-\sqrt{2\delta}\right] =0.
\end{equation}
Finding the conditions at which this cubic equation has a purely imaginary solution corresponding to the gap, we  can determine  the expressions for the maximal width of the secondary gap $\delta_c$ and the critical phase $\varphi_c$ at which the gap closes . Thus, we obtain for $E_{\textrm{Th}}\gg\Delta$
\begin{align*}
 \delta_c&\approx\left( 17/2-6\sqrt{2}\right) \left(\frac{\Delta}{E_{\textrm{Th}}}\right)^2,\\
 \varphi_c&\approx\sqrt{2\left( 5\sqrt{5}-11 \right)} \frac{\Delta}{E_{\textrm{Th}}},
\end{align*}
which is in agreement with our numerical results. Both the width and the critical phase $\varphi_c$ are small in this limit and the upper edge of the gap is attached to $E=\Delta$.
Additionally, from Eq.~(\ref{eq:curr_cons3}) the DOS in the region between both gaps can be calculated analytically and the result is shown in Fig.~\ref{fig:universal}. 
 For the upper minigap edge, we find
$ \delta_{\textrm{mini}}\approx\left( 17/2+6\sqrt{2}\right) (\Delta/E_{\textrm{Th}})^2
$.
The DOS above and below the gaps vanishes like square roots [$
N(\delta_{c}+x)/N_0\approx 2 (E_{\textrm{Th}}/\Delta)^2 \sqrt{x}/ (3 \sqrt{2}-4)^{3/2}$; 
$N(\delta_{mini}-x)/N_0\approx 2 (E_{\textrm{Th}}/\Delta)^2 \sqrt{x}/ (3 \sqrt{2}+4)^{3/2}$].
The maximum of the DOS in the region between the gaps lies at
$ \delta_{\textrm{max}}\approx (\Delta/E_{\textrm{Th}})^2/18$
and has the value $\sqrt{2} E_{\textrm{Th}}/\Delta$.

We attribute the origin of the secondary minigap to the level repulsion between the discrete levels with energy below $\Delta$ and the states of continuum having a divergent density of states at $E\to\Delta+0$. We may illustrate it by the solution of a model \cite{beenakker:92} containing a single resonant level instead of the cavity. It yields an Andreev level $E_A$ which never "sticks" to $E=\Delta$. In the limit of a broad resonance ($\Gamma\gg\Delta$) the corresponding minigap derived in the model of Ref. \cite{beenakker:92} is $\Delta-E_A\sim\Delta^3/\Gamma^2$. It is natural to associate qualitatively $E_{\textrm{Th}}$ of our problem with $\Gamma$, which indeed yields $\delta_\textrm{c}\sim (\Delta/E_{\textrm{Th}})^2$, in agreement with the rigorous result. Finding the numerical coefficient here and the phase dependence of the gap is beyond the scope of this illustration.

As already anticipated in Fig.~\ref{fig:setup}, the $E$-$\varphi$ plot of the secondary gap shows a smile shape, with a finite extension in phase and its maximum size in energy at zero phase difference. The dependence of the gap edges and $\varphi_c$ on $E_{\textrm{Th}}/\Delta\simeq 1 $ is displayed in the lower part of Fig.~\ref{fig:gaps} while   the shape of the gap in the $E$-$\varphi$ plane for a set of various Thouless energies  is shown in the upper part. We see that the secondary gap first increases with decreasing $E_{\textrm{Th}}$ yet reaches a maximum width $\simeq 0.01 \Delta$ at $E_{\textrm{Th}} \approx \Delta$ and further decreases. The upper gap edge detaches from $\Delta$ at $E_{\textrm{Th}}=\Delta$ and the secondary gap disappears entirely at  $E_{\textrm{Th}}\approx 0.68 \Delta$. The critical phase $\varphi_c$ exhibits a cusplike maximum at $E_{\textrm{Th}}$ slightly below the value at which the detachment takes place. The relatively small size of the secondary gap perhaps explains the fact that it has not been discovered in the previous numerical simulations. Furthermore, the phase dependence of the secondary gap also has interesting implications for the underlying Andreev level density. In the standard resonant level model \cite{beenakker:92}, Andreev levels always move toward the Fermi level with an increase in the phase difference. The opposite behavior of the levels at the lower edge of our secondary gap hints at the importance played by the energy-dependent correlations of the scattering channels for a finite Thouless energy.

For a general asymmetric setup with $G_1\neq G_2$, the phase $\phi$ on the node is no longer zero and current conservation provides two equations in two complex variables $\phi$ and $g$. Since the equations are straightforward but lengthy, we do not give them here. Asymmetry only enters our calculations via a dimensionless asymmetry parameter $a=G_1/G_2$. The dependence of the DOS on $a$ is shown in Fig.~\ref{fig:asymm} ($a$ and $1/a$ give identical pictures). 

We see that for asymmetric setups the situation becomes more complicated as more secondary gaps open in the DOS. It is worth noting that for $\varphi=0$ (left edge of each plot) the asymmetry does not manifest itself in the DOS, since such a setup is equivalent to a single superconductor connected to a normal metal with a single contact with total conductance $G_1 +G_2$ . With increasing asymmetry, the effect of the superconductors is dominated by the stronger contact, since the phase on the node becomes "pinned" to the phase of the more strongly coupled superconductor. Thus, the overall phase dependence of the DOS gets weaker and approaches the $\varphi=0$ result for almost all phases and energies. Despite this, 
qualitative changes occur at energies close to $\Delta$ (upper row of the plots).
There, at increasing $a$ we observe a formation of yet another gap centered at $\varphi = \pi$. In the limit of strong asymmetry, both gaps fill almost all space above a certain energy. However, they are always separated by a thin strip of finite DOS. 
In the lower row of plots, we concentrate at energies close to zero. There, we see the usual minigap $\simeq E_{\textrm{Th}}$ with the lower edge attached to zero. As known \cite{zhou:98}, the usual minigap closes at $\varphi=\pi$. A new element is yet another secondary gap emerging at finite $a$ around $\varphi=\pi$. Similar to the upper row, we see that this gap is also complementary to that centered at $\varphi=0$. In the limit of large asymmetry both gaps fill almost all space below a certain energy. However, as in the upper row of plots, they are always separated by a thin strip of finite DOS.  

\begin{widetext}
\begin{center}
\begin{figure}[thb] 
 \includegraphics[width=0.9\columnwidth,angle=0,clip=true]{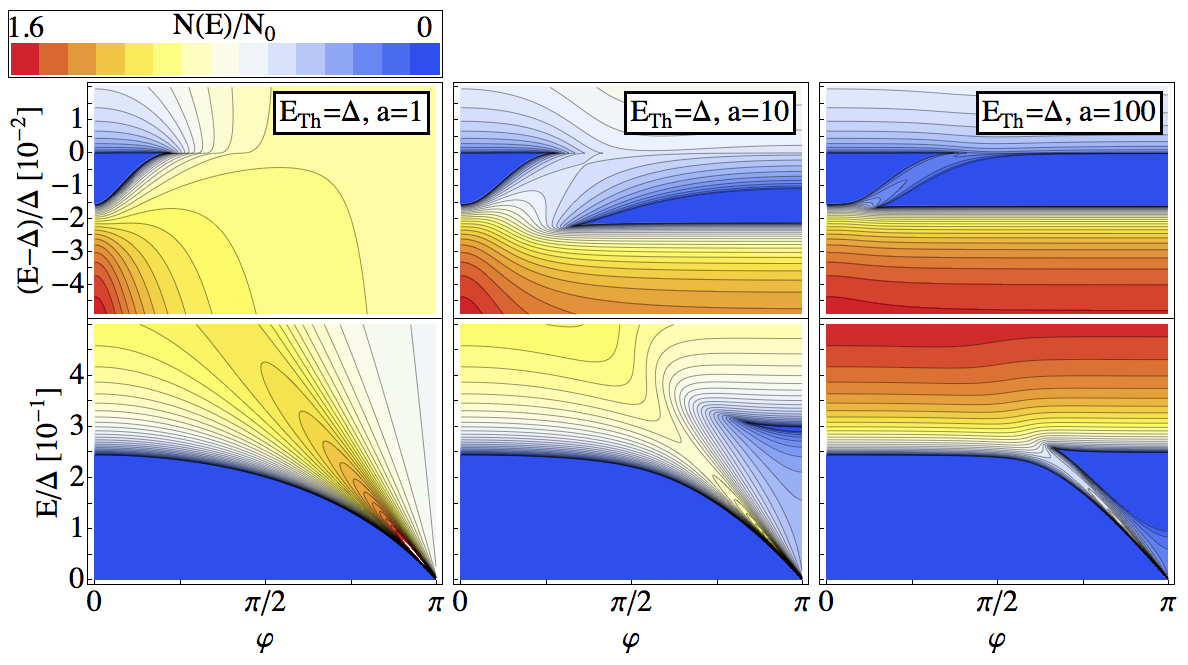}
\caption{\label{fig:asymm}
Dependence of the local density of states on increasing asymmetry $a=G_1/G_2$. The lower part of each plot shows the well-known minigap, the upper part shows the secondary gap. In each plot $E_{\textrm{Th}}=\Delta$. For a symmetric setup (left plot) the result is equivalent to Fig.~\ref{fig:setup}. With increasing asymmetry $a=10$ (central plot) and $a=100$ (right plot) both gaps are stable, however, additional gaps appear around $\varphi=\pi$.}
\end{figure}
\end{center}
\end{widetext}

The secondary gaps we found for the chaotic cavities also persist for more general contacts, as long as the distribution of the transmission eigenvalues has a gap between  $T=0$ and some finite $T_{\textrm{min}}$ value. Conversely, the gaps do not appear if a substantial fraction of the transmission eigenvalues is close to zero, as is the case for tunnel, diffusive \cite{dorokhov:82}, or dirty contacts \cite{qt}. We note that in our Letter, the so-called Ehrenfest time separating semiclassical from wave dynamics is small and plays no role, different from other studies \cite{beenakker:05,kuipers:10}. Furthermore, our predicted secondary gaps are also robust against a weak spatial dependence as we discuss in the Supplemental Material \cite{suppl}. 


To conclude, we have shown that a smile-shaped secondary gap just below the superconducting gap edge $\Delta$ appears in the density of states of a cavity between two superconductors. The gap becomes small for large Thouless energies, closes at a finite phase difference between the superconductors, and disappears at a critical $E_{\textrm{Th}} \simeq 0.68 \Delta$. These gap features are robust against asymmetries of the contact conductances and nonballistic contacts involving transmissions smaller than one. For an asymmetric setup, we have found two more additional gaps centered at phase difference $\pi$. It would be interesting to experimentally observe our predictions, e.g., in multiterminal semiconductor or carbon nanotube cavities by means of tunneling spectroscopy. On the theoretical side, it presents a challenge to explore in more detail the level correlations at the critical points, when the secondary gap closes with phase.

J.~R. and W.~B. were supported by the DFG through SFB 767 and Grants No. BE 3803/5 and by the Carl Zeiss Foundation. Y.~N. and L.~G. thank  the Aspen Center for Physics, supported in part by NSF Grant No. PHYS-1066293, for hospitality. Work at Yale is supported by NSF DMR Grant No. 1206612.


\end{document}